\documentclass{cjaa}                    

\input{psfig.sty}                       

\begin{document}
   \title{ 1RXS J232953.9+062814: a New SU UMa Dwarf Nova below the Period Minimum}

   \author{Jian-Yan Wei
      \inst{}\mailto{}
   \and Xiao-Jun Jiang
      \inst{}
   \and Da-Wei Xu
     \inst{}
   \and Ai-Ying Zhou
     \inst{}
   \and Jing-Yao Hu
     \inst{}
      }
   \offprints{Jian-Yan Wei}                   

   \institute{National Astronomical Observatories, Chinese Academy of Sciences,
             Beijing 100012, China\\
             \email{wjy@bao.ac.cn}
          }

   \date{Received~~2001~~~~~~~~~~~~~~~ ; accepted~~2001~~~~~~~~~~~~~~ }

   \abstract{
1RXS J232953.9+062814 was identified as a cataclysmic variable
by Wei et al. (1999). Four low-resolution spectra of 1RXS J232953.9+062814
were obtained by using the 2.16-m telescope of the National Astronomical
Observatories, in which two of them were at outburst, and the other two
were at quiescence. The system is about 16.8\,$B$ and 16.5\,$V$ at quiescence, and
12.6\,$B$ and 12.6\,$V$ at outburst. The quiescent spectra were dominated by
double-peaked Balmer emissions, which indicates a hydrogen-rich system
with a high-inclination accretion disc. MgH and TiO absorption bands
appeared in the quiescent spectrum imply a companion with a spectral type
of early M dwarf. If we take it as a M0 dwarf, the system is located at a distance of 350\,pc
with a proper motion velocity 150\,km s$^{-1}$. The superhump
period of 0.046311\,days (Uemura et al. 2001) was confirmed by our $V$ photometry.
The short period and the hydrogen-rich nature reveal that this system is
another SU Ursae Majoris-type dwarf nova below the period
minimum after V485 Centauri. 1RXS J232953.9+062814 is one of the most important
systems for studying the evolutionary scenario of cataclysmic variables
since it is much brighter than V485 Cen.
   }

   \authorrunning{J.-Y. Wei et al.}
   \titlerunning{1RXS J232953.9+062814: a New SU UMa Dwarf Nova below the
Period Minimum }

   \maketitle
%
%
\section{Introduction}              
\label{sect:intro}
1RXS J232953.9+062814 was identified as a cataclysmic variable (CV) by
Wei et al.\ (1999) when they selected a bright AGN sample from the optical
identifications of the Bright Source Catalog of ROSAT All Sky
Survey (RASS-BSC) (Voges et al.\ 1999). It was classified as a dwarf nova
by Hu et al.\ (1998).
The CCD photometric observations by Uemura et al.\ (2001) on
2001 Nov. 4.47--6.17 revealed superhumps with amplitudes of 0.2--0.3\,mag and
a period of 0.046311(12) days, indicating that this object is an SU UMa-type
dwarf nova. This short superhump period means that the orbital period of
this object is below the `period minimum' (about 1.3\,h). Our low-resolution
quiescent spectra shows it is a hydrogen-rich system. Except for this
object, the only known object in this class is V485 Cen (Augusteijn 
et al.\ 1996, Olech 1997 ).
However, 1RXS J232953.9+062814 is much brighter, hence much easier
to be observed than
V485 Cen ( about 18.0\,$V$ at outburst ). Uemura et al.\ (2001) also pointed out that
this object shows large proper motion (up to 0\arcsec.1 yr$^{-1}$),
which indicates a small distance for the system. Thus this object is one
of the most important systems for studying the evolutionary scenario of
cataclysmic variables.

We observed four times on 1RXS J232953.9+062814 spectroscopically after
its discovery on 1996 Nov. 5 to 2001 Nov. 8. It was in outburst by chance
when it was observed for the first time on 1996 Nov. 5.
In Wei et al.\ (1999) only the quiescent spectrum obtained on 1997 June 6
was published. In this paper we publish all the spectra and a preliminary
results of new photometric observations,
paying special attention to the hydrogen-rich nature of this system
and the spectral classification of the companion, which is very important
to understand the nature of this object.

\section{Observations }
\label{sect:Obs}
A star on the Digitized Sky Survey (DSS) with
RA=$23^{\rm h}29^{\rm m}54^{\rm s}$.38 and Dec=$+06^{\circ}28'10^{\prime\prime}$.2 (2000.0)
was identified as the optical counterpart for 1RXS J232953.9+062814 (Wei et al. 1999).
The spectrum of the optical counterpart was observed by using the 2.16-m telescope of
the National Astronomical Observatories of CAS (NAOC) four times.
The observations were carried out on 1996 Nov. 5, 1997 June 10 and 11,
and 2001 Nov. 8, respectively. All the observations were done with
the OMR low-resolution Cassegrain spectrograph equipped with
a TEK 1024$\times$1024 back illuminated CCD camera with pixel size of 24 microns .
The spectra ranged from 3800 to 8200\,\AA~with a resolution $\sim$9\,\AA.
He/Ar lamp was taken as the wavelength calibration, and
2 or 3 KPNO standard stars were observed every night for flux calibrations.
All the data were reduced with IRAF using standard method.

In addition, immediately after the detection of the outburst of the dwarf nova
on 2001 November 3.926 UT (Uemura et al.\ 2001),
Johnson $V$ photometry covering four periods was performed with the three-channel
high-speed photoelectric photometer (Jiang \& Hu 1998) on the 85-cm Cassegrain
telescope of NAOC on 2001 November 8.

\section{Results}
\label{sect:analysis}
All of the four spectra are presented in Fig.~1. The CV was in outburst
on 1996 Nov. 5, while on 2001 Nov. 8, the CV was on the midway in fading,
the spectrum shows
emission features on blue continuum. And the quiescent spectrum obtained
on 1997 June 10 had a low signal to noise ratio due to the cloudy sky during
the observations. The spectrum in outburst (panel `a' in Fig. 1) clearly
showed broad Balmer absorption lines,
which liked the spectrum of a typical DA white dwarf.
And the two quiescent spectra presented Balmer and He\,I emission lines,
which indicates that 1RXS J232953.9+062814 is not an AM CVn star but a
hydrogen-rich system. This hydrogen-rich nature and the 0.046311(12) days
superhump period indicate that this object is certainly below the `period
minimum'. The emission lines of the second quiescent spectrum also showed
double-peaked emission features (send spectrum requests to: Jian-Yan Wei),
which indicates a high-inclination accretion disc.

For this system, 16.6\,$B$ and 15.6\,$R$ were given by USNO A1.0( Monet et al.\ 1996 ).
The $B$ and $V$ magnitudes were estimated from the spectra showed in Fig.~1.
They are listed as the following: 12.6\,$B$ and 12.6\,$V$, 1996 Nov. 5; 17.1\,$B$ and 16.5\,$V$,
1997 June 11; 14.9\,$B$ and 14.7\,$V$, 2001 Nov. 8. They are coincident with the mean
magnitude of 16.8\,$B$ in quiescence obtained by Zharikov \& Tovmassian (2001, VSNET
homepage) on 2001 October 25 and 26, and the brightness of 12.5\,$V$ detected
by Uemura et al.\ (1997) on 2001 Nov. 3.926\,UT during its outburst.
The outburst on 1996 Nov. 5  had the similar amplitude of about 4\,mag.

The quiescent spectrum of 1997 June 11 showed strong stellar absorption
features from the companion. We can check out the MgH features at
4780 and 5211\,\AA, and TiO bands around 6159 and 6178\,\AA.
These MgH and TiO features indicate that the companion has a spectrum of
early M-type. The brightness of the companion should be
close to 16.5\,$V$ since the flux from
the companion dominates the continuum of the quiescent spectrum at 5500\,\AA.
The system locates at a distance of $\sim$350\,pc, and has a
proper motion velocity of 150\,km s$^{-1}$, if we consider the companion is a M0 dwarf.

A Fourier analysis based on our $V$ photometric data resulted in a 
period of 0.046\,d and
a poor fitting residual of $\sim$0.26\,mag compared to the 
observational scatter.
We fitted the light curves assuming a period of 0.04633\,d for superhump
(see Fig.~2), and obtained a $V$ amplitude of 0.351$\pm$0.008\,mag 
for the outburst.
The observations probably covered just after the 
rapid decline phase,
further investigation on available light curves is obviously needed.
%

\begin{figure*}[hbtp]
\psfig{figure=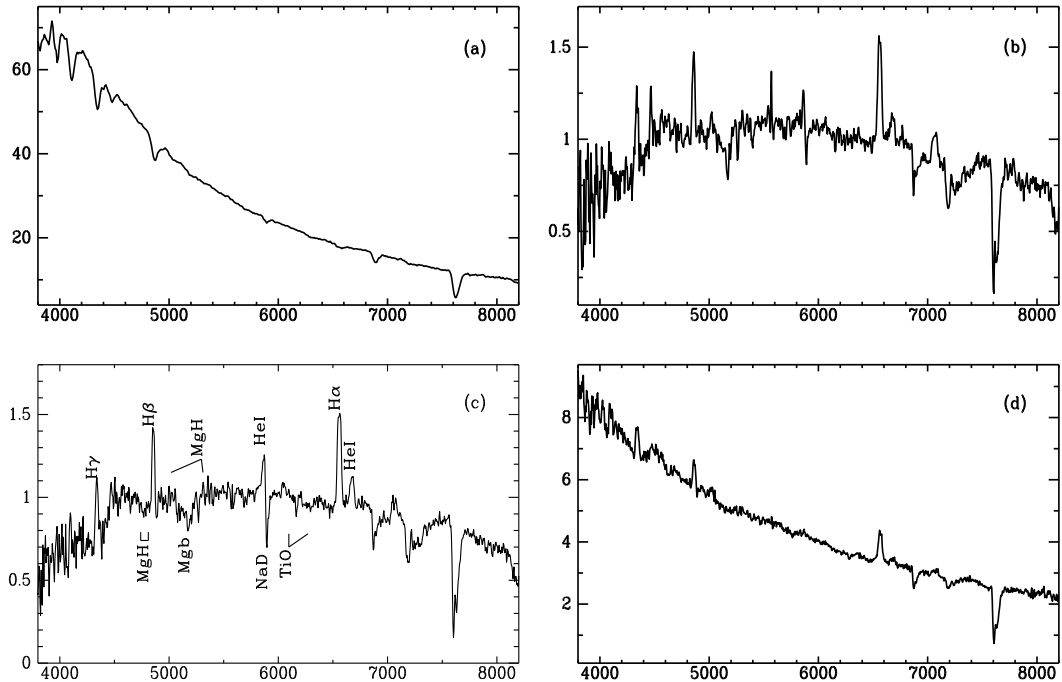,width=140mm,height=90mm}
        \caption{Optical spectra of 1RXS J232953.9+062814.
        f$_\lambda$ in units of 10$^{-15}$~erg~cm$^{-2}$~s$^{-1}$~\AA$^{-1}$
        is plotted against wavelength in \AA. Panel a--d were taken on
1996 Nov. 5,1997 June 10, 1997 June 11, and 2001 Nov. 8, respectively.}
\end{figure*}

\begin{figure}
   \begin{center}
   \vspace{2mm}
   \hspace{3mm}\psfig{figure=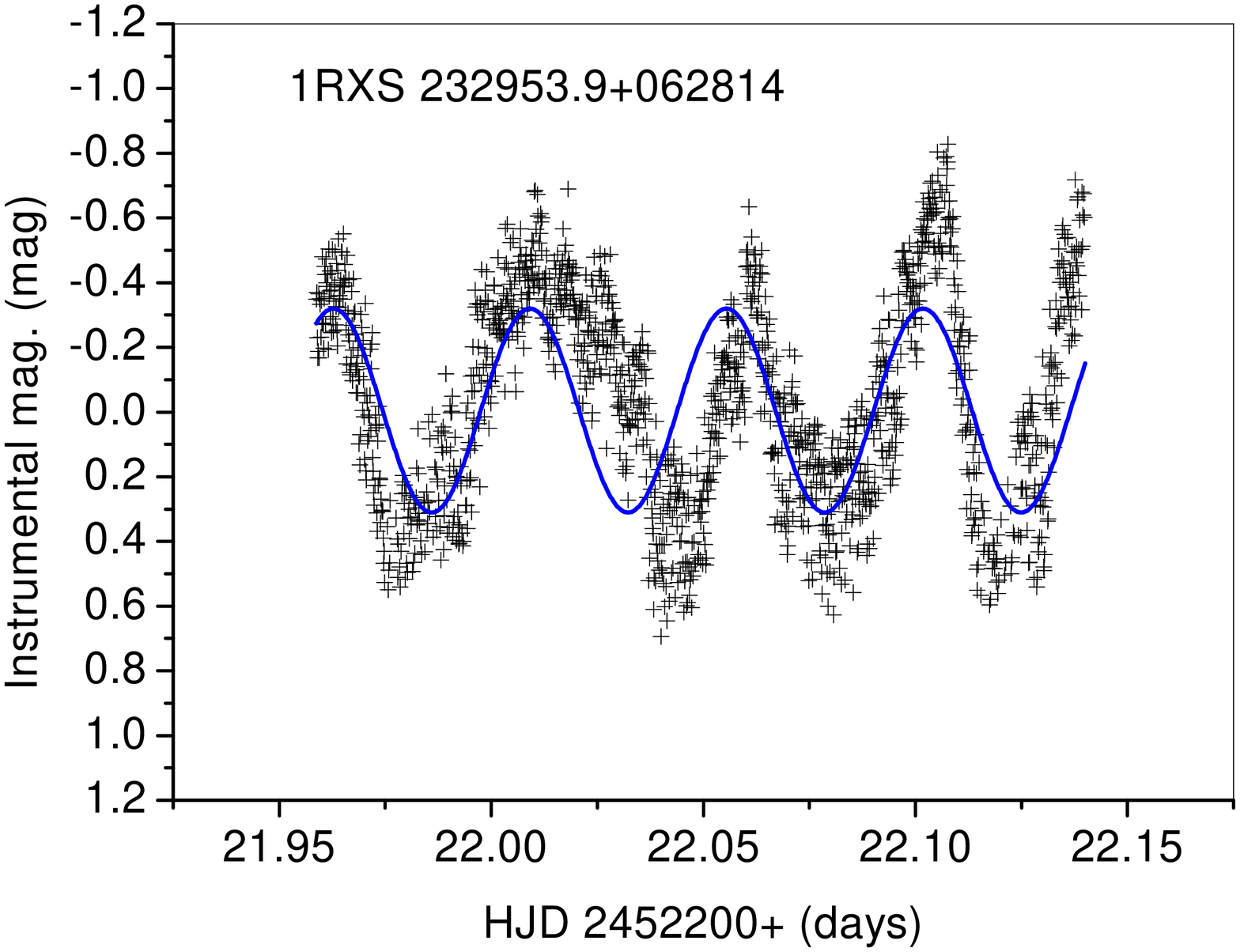,width=140mm,height=120mm,angle=0.0}
   \parbox{180mm}{{\vspace{2mm} }}
   \caption{ $V$ light curves (crosses) of 1RXS J232953.9+062814 during its superoutburst
   on 2001 November 8 together with a sinusoid with period of 0.04633\,d.
   The integration time for each point is 10\,s.  }
   \label{Fig:lightcurve-J2329}
   \end{center}
\end{figure}

\begin{acknowledgements}
We would like to thank Prof. Huilai Cao for donating telescope time, we also
thank Mr. Hongbin Li for his assistance in observations. This work was funded by
the Natural Science Foundation of China.
\end{acknowledgements}

\label{lastpage}

\end{document}